\documentclass[11pt,onecolumn,amsmath,amssymb,nofootinbib,AMS]{revtex4}

\usepackage{dcolumn}
\usepackage{bm}
\usepackage[latin1]{inputenc}
\usepackage[spanish,english]{babel}
\usepackage{amsfonts}
\usepackage{amssymb}
\usepackage{graphicx} 
\usepackage{color}

\newcommand{\be}{\begin{equation}}
\newcommand{\ee}{\end{equation}}
\newcommand{\bea}{\begin{eqnarray}}
\newcommand{\eea}{\end{eqnarray}}

\begin{document}

\title{Creation of quantized particles, gravitons, and scalar perturbations by the expanding universe}
\author{Leonard Parker}
\affiliation{Center for Gravitation, Cosmology and Astrophysics, Physics Department, University of Wisconsin-Milwaukee, USA}

\begin{abstract} 
{
Quantum creation processes during the very rapid early expansion of the universe are believed to give rise to temperature anisotropies and polarization patterns in the CMB radiation. These have been observed by satellites such as COBE, WMAP, and PLANCK, and by bolometric instruments placed near the South Pole by the BICEP collaborations. The expected temperature anisotropies are well-confirmed.  The B-mode polarization patterns in the CMB are currently under measurement jointly by the PLANCK and BICEP groups to determine the extent to which the B-modes can be attributed to gravitational waves from the creation of gravitons in the earliest universe. 

As the original discoverer of the quantum phenomenon of particle creation from vacuum by the expansion of the universe, I will explain how the discovery came about and how it relates to the current observations. The first system that I considered when I started my Ph.D. thesis in 1962 was the quantized minimally-coupled scalar field in an expanding FLRW (Friedmann, Lemaitr{\' e}, Robertson, Walker) universe having a general continuous scale factor $a(t)$ with continuous time derivatives. I also considered quantized fermion fields of spin-1/2 and the spin-1 massless photon field, as well as the quantized conformally-invariant field equations of arbitrary integer and half-integer spins that had been written down in the classical context for general gravitational metrics by Penrose.

It was during 1962 that I proved that quanta of the minimally-coupled scalar field were created by the general expanding FLRW universe. This was relevant also to the creation of quantized perturbations of the gravitational field, since these perturbations satisfied linear field equations that could be quantized in the same way as the minimally-coupled scalar field equation. In fact, in 1946, E.M. Lifshitz had considered the classical Einstein gravitational field in FLRW expanding universes and had shown that the classical linearized Einstein field equations reduced, in what is now known as the Lifshitz gauge, to two separate classical minimally-coupled massless scalar field equations. These field equations of Lifshitz, when quantized, correspond to the field equations for massless gravitons, one equation for each of the two independent polarization components of the spin-2 massless graviton. I will discuss this further in this article.\footnote{Plenary Lecture given September 2, 2014 at the ERE2014 Conference in Valencia, Spain.}

}
\end{abstract}

\maketitle

\section{Introduction}
\label{Introduction}

{\textbf 
In 1962, at Harvard when I began my Ph.D.~thesis, I wanted to work at the interface of general relativity and quantum field theory.  I had already studied general relativity quite carefully as an undergraduate before coming to Harvard.  I had the good fortune to 
learn quantum field theory and particle physics at Harvard from Wendell Furry, Roy Glauber (Nobel Prize 2005), Sidney Coleman, Sheldon Glashow (Nobel Prize 1979), and Julian Schwinger (Nobel Prize 1965). 

I wanted to find new consequences of the quantum field theory of elementary particles in the context of Einstein's theory of general relativity.  At the time, I felt that {\em quantizing} the nonlinear gravitational field {\em itself} was so difficult that I would not be able to make significant progress in trying to go beyond the deep work that had already been done in that area.  Nevertheless, I felt that it would be valuable to study quantized elementary particle fields in the curved space-times that were solutions of the classical nonlinear Einstein gravitational field equations. Luckily, Sidney Coleman agreed to be my thesis advisor on such a project, which was outside the main stream of the time.

I started by looking for new consequences of quantum field theory in the isotropically expanding cosmological space-times that were solutions of the nonlinear classical equations of general relativity.
To investigate the consequences of quantum field theory in the expanding universes of general relativity, it was first necessary to extend the known quantum field theory of elementary particles from the well-established flat Minkowski space-time of special relativity to the context of a classical general relativistic isotropically expanding universe with a general expansion scale-factor, $a(t)$. Some examples that I studied were the dust-filled universe, the radiation-dominated universe and the exponentially-expanding universe of deSitter, but my main result was for general $a(t)$.

The quanta (particles, scalar perturbations, and tensor gravitational wave perturbations) that reach a steady-state density during the inflationary stage of the expansion are created or maintained by the quantum field theory process of particle creation that I discovered and will discuss in this lecture. The quantum field theory vacuum state in the exponentially expanding universe is known as the Bunch-Davies\cite{Bunch-Davies1979} vacuum.  That vacuum state is symmetric under the deSitter invariance group of the exponentially expanding FLRW universe.  The Bunch-Davies vacuum is the natural vacuum state having the 10 deSitter group symmetries\cite{Higuchi1987, Parker-Toms2009}, just as the natural vacuum state in flat Minkowski spacetime has the 10 Poincar{\'e} group symmetries.

The constant, non-zero, average number density of particles, scalar perturbations and/or tensor gravitational wave perturbations is consistent with the result of spontaneous particle creation from vacuum of an FLRW universe that undergoes the deSitter exponential expansion. I will discuss this further in Section \ref{exponential expansion}, and will show that the creation rate of these perturbations and/or particles follows immediately from the result that I obtained in my Ph.D. thesis for general continuous expansion factors $a(t)$ having a sufficient number of continuous time derivatives. This follows very simply from my {\em general} equation for the average number of particles created, and from the deSitter invariance of the exponentially expanding universe.

I chose to first consider the spatially-flat FLRW universes because I felt that one could {\em dynamically propagate} the experimentally verified theory of quantized fields from flat spacetime into the curved spacetime of the spatially-flat isotropically expanding universes. It was not necessary in this framework to go beyond the already tested assumptions of classical general relativity and of quantum field theory in {\em flat} spacetime. The dynamical equations of the quantized fields were taken to be the simplest generally-covariant ones obtained from the flat-spacetime quantized field equations. It was then possible to use the field equations in the curved spacetime to propagate the fields and their commutation relations into curved spacetime. Any fundamentally new results coming from this minimal set of assumptions would have to be taken very seriously. As it turned out, this procedure also gave an independent new proof of the spin-statistics theorem; for bosons only the commutation relations were propagated consistently from the initial to the final flat spacetime, and for fermions only the anticommutation relations were propagated consistently. 

First, I considered {\em the quantized generally-covariant minimally-coupled free scalar-field equation} in an isotropically-expanding spatially-flat universe with a continuous and smooth scale factor $a(t)$. Because there was no ambiguity as to the definition of the measurable particle number operator of the quantized field in flat spacetime, it was natural to consider what happened to the particle number when the scale factor $a(t)$ went smoothly from an initial constant value $a_1$ at early times to a final constant value $a_2$ at late times. During the periods of time when $a_1$ and $a_2$ were constant, simple constant rescalings of the spatial coordinates at early and late times showed that spacetimes before and after the expansion of the universe were Minkowskian, each exhibiting the 10-dimensional Poincar\'{e} group of symmetries. The number density of particles of the quantized field was taken to be $0$ at early times. Evaluating the expectation value of the number density at late times, after the universe had undergone a smooth expansion, I found that the number density of created particles was finite and nonzero!  This demonstrated that real particles were created from the initial vacuum state by the expansion of the universe.

To briefly summarize the calculation, I used the Heisenberg picture to evolve the quantized field during the expansion of the universe. The state vector of the quantized field was chosen to be the well-defined vacuum state in the initial Minkowski spacetime. As a result of the time-dependent unitary evolution of the quantized field, the creation and annihilation operators of the quantized field at {\em early} times had evolved into linear combinations of the creation and annihilation operators of the quantized field at {\em late} times. Consequently, the state vector having no particles of the quantized field at early times does have a non-zero density of particles of the quantized field at late times. This implies that particles are created by the expansion of the universe. 

My results first appeared in my Ph.D. Thesis\cite{Parker1966} (``The Creation of Particles by the Expansion of the Universe", Harvard University, 1966) and also in \cite{Parker1968, Parker1969, Parker1971}. My thesis also included results I had obtained for the spin-1/2 fermion field, including particle creation for fermions of non-zero mass, as well as the absence of particle creation for the {\em massless} Pauli neutrino field. Those results for fermion fields appeared in \cite{Parker1971}.
  
I also found that particles of {\em $0$-mass} satisfying the minimally-coupled scalar field equation {\em would be created} from the initial vacuum state as a result of the expansion from an initial Minkowski spacetime to a final Minkowski spacetime. This result implied that quantum gravitons would be created by the quantized linearized Einstein gravitational field equations.  E.M. Lifshitz\cite{Lifshitz1946} in 1946 had published an important paper on the {\em classical} linearized Einstein gravitational field equations, in which he proved that there is a gauge (the Lifshitz gauge) in which the linearized Einstein field equations reduce to a set of two separate minimally-coupled scalar field equations. One could then use the Lifshitz gauge to directly apply my results to the Einstein gravitational field to show that quantized gravitons (tensor perturbations) were created by the expanding universe.  A paper on this was published by Leonid P. Grischuk\cite{Grishchuk1975}, in which he used the Lifshitz gauge to apply my earlier quantization of a minimally-coupled scalar-field, including my expression for particle creation in a general expanding universe. He used the Lifshitz gauge to obtain an expression for the creation of gravitons in a universe with $a(t)$ proportional to a power of $t$.

A more detailed quantization of the linearized Einstein gravitational field was published in 1977 by Lawrence H. Ford and me in \cite{Ford-Parker1977}. One of the things we showed in our paper was that there were spurious infrared (IR) divergences present in Grishchuk's result that were not present when we imposed more appropriate late-time boundary conditions.  In our paper, we also reviewed the derivation of the Lifshitz gauge and performed an explicit {\em quantization}, defining creation and annihilation operators for massless gravitons in each of their two polarization states.  (We carried this out for more than one way of specifying the pair of gravitational wave polarization states.)

The minimally-coupled scalar field equations that appear in the Lifshitz gauge are the same ones I studied \cite{Parker1966, Parker1968, Parker1969} in the 1960's for general $a(t)$. In the exponentially expanding inflationary universe, there are well-known analytic solutions of these equations that can be applied to the creation of gravitons and gravitational wave perturbations present in the inflationary universe. These form the basis for obtaining the inflationary universe predictions of linearized gravitational waves and gravitons created in the early universe, as well for scalar gravitational perturbations that are related to the temperature variations of the CMB.

In Section \ref{exponential expansion}, I will show how one obtains directly from my published results of the 1960's, the average number of quantized particles created as a function of time in an expanding physical volume of the exponentially expanding universe. This result follows directly from deSitter invariance of the exponentially expanding universe. In addition, if there are similar particles already present at the time that the exponential expansion begins, then the initial presence of these particles can significantly increase the rate of particle creation. This stimulated creation of identical bosonic particles occurs also for general $a(t)$. (For fermions the initial presence of identical fermions would decrease the rate of particle creation.)

Scalar perturbations of the gravitational field were also created by the exponential expansion of the universe, before reheating occurred. Reheating resulted in the formation of a plasma of charged particles in a high- temperature, radiation-dominated expanding universe. The acoustic waves that these scalar gravitational perturbations produced in the plasma were present for about 400,000 years, until the universe cooled enough for electrons and protons to form neutral atoms, after which the photons of the CMB were able to propagate freely (for the most part). However, temperature differences in the plasma acoustic waves present at the time of last scattering resulted in temperature differences in the free-streaming CMB radiation. These temperature differences in the CMB radiation at the time of last scattering were responsible for the temperature anisotropies of about 1 part in $10^5$ that were ultimately measured by the COBE satellite.

If the recent possible gravitational wave observations are confirmed and if the intensities of the B-mode polarization patterns they produce are consistent with the inflationary universe model, including the observed temperature anisotropies in the CMB, then this consistency would be further evidence for an early inflationary expansion of the universe.

In the next Section, I will briefly explain how I proved in my Ph.D. thesis\cite{Parker1966}  (see also \cite{Parker1968, Parker1969, Parker1971}) that minimally-coupled scalar particles are created by expanding FLRW universes.

But first let me quote the Conclusion of my 1968 paper\cite{Parker1968}:
``The particle creation in the expanding universe at the present time is quite negligible. However, for the early stages of a Friedmann expansion it may well be of great cosmological significance, especially since it seems inescapable if one accepts quantum field theory and general relativity. In considering the large amount of particle creation taking place in the early stages of an expansion, it is necessary to take into account the reaction of the matter created back on the gravitational field. Furthermore, it may be necessary to consider the effects of the quantization of the gravitational field. Therefore, no conclusive quantitative result can yet be reported here concerning the primeval creation."

This conclusion of my paper emphasized the strong effects that particle creation would have on the early expansion of the universe, so the reaction-back would have to be taken into account. The first such calculation was done by me and Steven A. Fulling \cite{Parker-Fulling1973}, in which we showed by numerical evolution of the semi-classical Einstein equations that a bounce could occur as a result of the particle creation process, thus avoiding the cosmological singularity.  Our result also demonstrated that the classical energy-conditions can be avoided through the creation of particles by quantum field theory in an FLRW universe. 

In my thesis \cite{Parker1966}, I also used quantum measurement theory to develop the method of adiabatic regularization (or renormalization) as applied to the particle number operator during the expansion of the universe. In 1974, Fulling and I extended the adiabatic regularization method to renormalize the expectation value of the energy-momentum tensor \cite{Parker-Fulling1974, Anderson-Parker1987, Parker-Toms2009}.

An interesting sidelight, is that I was hired as an Instructor to teach at the University of North Carolina (UNC) for 1966 and 1967. My thesis advisor was away in Europe, so I had to wait until he returned to Harvard in 1966 to fly to Boston for my thesis defense. The thesis defense committee consisted of Sidney Coleman, Sheldon Glashow, and Walter Gilbert (Nobel Prize 1980). 

While teaching at UNC, I attended a Colloquium given at UNC in 1966 or 1967 by Fred Hoyle.  After his talk, I explained to him my as yet unpublished work.  He was an advocate of the steady state theory (of Bondi, Gold, and Hoyle), which assumed a permanent exponentially expanding universe having a constant density of particles, maintained by a process that they had postulated, in which particles or atoms were constantly being created, including at the present time, at a sufficient rate to maintain a steady-state universe that did not change significantly on cosmological time scales.  I explained to him that my quantum field theory mechanism of particle creation would not support such a steady-state universe at the present time because the particle creation rate would be much too small. However, I explained to him that my process of quantum field theory particle creation could support an exponentially expanding steady-state universe at very early times with a much higher particle density. He listened, but was not enthusiastic about the idea because at that time he believed in a universe that was eternal and unchanging.

In Section \ref{exponential expansion} of this article, I will show how the equations that I first derived in my Ph.D.thesis\cite{Parker1966, Parker1968, Parker1969, Parker1971} for the general case of particle creation by an FLRW universe directly yields the rate per unit physical volume at which particles or quanta are created in an exponentially expanding, spatially-flat FLRW universe that is in (or has evolved to) a deSitter invariant state. For example, the inflationary universe is in such a state prior to reheating.
This very high rate of particle creation is what I was referring to in connection with the exponentially expanding steady-state universe in the discussion I had with Fred Hoyle after his Colloquium at UNC.
}

\section{Minimally-coupled scalar field in FLRW universe}
\label{sec-scalar}

{\textbf
Let us now consider the case of a free quantized scalar field $\phi$ in a smoothly changing spatially flat FLRW universe with line element
\be 
ds^2 = dt^2 - a^2(t)(dx^2 + dy^2 + dz^2), \label{2-1}
\ee
and equation of motion
\be 
(\Box + m^2)\phi = 0, \label{2-2}
\ee
where $\Box= g^{\mu\nu} \nabla_\mu \nabla_\nu$, with $\nabla_\lambda$ denoting the covariant derivative. 

For now 
we are considering the so-called minimally-coupled scalar field of mass $m$ (which may be $0$).  
More generally, one could include a coupling to the Ricci scalar curvature,
$R=g^{\mu\nu} R_{\mu\nu}$, namely, $(\Box + m^2 + \xi R)\phi = 0$, where $\xi$ is 
a dimensionless constant. We work in units with $\hbar$ and $c$ each equal to $1$.  
(For brevity, I will not go into spin-1/2 and spin-1 fields, and spatially-curved FLRW universes, 
which I also considered in my Ph.D. thesis\cite{Parker1966, Parker1968, Parker1969, Parker1971}.

Writing  (\ref{2-2}) with the metric of  (\ref{2-1}), one finds that
\be 
a^{-3} \partial _t(a^3\partial _t\phi ) - a^{-2} \sum _i
\partial^2_i\phi + m^2 \phi = 0. \label{2-3}
\ee
It is convenient to impose periodic boundary conditions in a cube having
sides of coordinate length $L$ and coordinate volume $V = L^3$.  
The physical volume of this cube is expanding, with the physical volume proportional to $a(t)^3$.  
As in Minkowski spacetime, this is a mathematical device, with $L$ taken to infinity
after physical quantities such as the density of created particles have been calculated.  
Then we can expand the field operator $\phi$ in the form
\be
\phi = \sum _{\vec k}\left\{ A_{\vec k} f_{\vec k}(\vec x, t)
+ A^\dagger_{\vec k} f^*_{\vec k}(\vec x, t)\right\}, \label{2-4a}
\ee
where
\be 
f_{\vec k} = (2V a(t)^3)^{-1/2} e^{i\vec k\cdot\vec x} h_k(t). \label{2-4b}
\ee
Here $k^i = 2\pi n^i/L$ with $n^i$ an integer, $k = \vert\vec k\vert$, and $h_k(t)$ satisfies
\be
 {d^2\over{dt^2}} h_k(t) +  {k^2\over a^{2}} h_k(t) + m^2 h_k(t ) - {3\over 4} \left({\dot a \over a}\right)^2 h_k(t)
  - {3\over 2} {\ddot a \over a} h_k(t) = 0. \label{2-5}
\ee

Examples in which this equation has analytic solutions are power law expansions and the exponential expansion. For the latter case, $a(t) \propto \exp(H t)$ with H being the Hubble constant during the exponential inflationary stage of the expansion, the solutions are Hankel functions. The metric has the 10 symmetries of deSitter spacetime, and a solution having these symmetries is
\be
f_{\vec k} = (2V a(t)^3)^{-1/2} e^{i\vec k\cdot\vec x} H_{\nu}^{1} (v). \label{2-4c}
\ee
with 
\be
v = k H^{-1} \exp(-H t)
\ee
\be
\nu = (9/4 - m^2/H^2)^{1/2}
\ee
Note that $v$ decreases with increasing $t$. Alternatively, one could replace $H_{\nu}^{1} (v)$ in (\ref{2-4c}) by $H_{\nu}^{2} (u)$ with $u = -v$, where the variable $u$ increases with $t$. (See \cite{Parker-Toms2009}, Sec.2.10, for a discussion of deSitter spacetime and the Bunch-Davies vacuum state.)

As already mentioned in Section \ref{Introduction}, the linearized Einstein equations for a gravitational wave field in the Lifshitz gauge reduce to a pair of massless minimally-coupled scalar field equations, one equation for each polarization of the gravitational wave. It is now clear that quanta of the gravitational wave field, i.e., gravitons, would be created by the expansion of the universe, in the same way that massless, minimally-coupled scalar particles would be created. For detailed quantization of linearized gravitational waves, see \cite{Ford-Parker1977}.

Now let us continue with our discussion of the minimally-coupled scalar field to summarize the method that I used to arrive at the number density of particles created from the initial Minkowski vacuum state (i.e., spontaneous particle creation) and from a non-vacuum state (i.e., stimulated particle creation).  Since the reaction back of the created particles would be very significant in the early expansion of the universe, I chose to deal with an arbitrary expansion of the universe that began in a Minkowski spacetime and ended in a Minkowski spacetime (to within a constant rescaling of the coordinates) and that had continuous functions $a(t)$, $\dot{a}(t)$, and $\ddot{a}(t)$. 

From the theory of ordinary differential equations there are two linearly independent solutions of (\ref{2-5}). One can prove that the particle number at late times is an adiabatic invariant and that the solutions that approach positive and negative frequency solutions in the late time Minkowski spacetime approach the asymptotic forms of the Liouville adiabatic solutions of the harmonic oscillator equation with slowly changing frequency. (This is also the form of the JWKB solutions for a slowly changing quantum mechanical scattering potential in one spatial dimension, with the particle energy above that of the scattering potential - but in the present case, the spatial coordinate $x$ is replaced by the time coordinate $t$.)

The adiabatic approximation of Liouville for the harmonic oscillator with a slowly changing frequency gives the following approximation to the solution of (\ref{2-5}):
\be
h_k(t )\sim (\omega_{k}(t))^{-1/2} \exp (\pm i \int^t \omega_{k}(t') \,dt'), 
\label{2-7}
\ee
where 
\be
\omega_{k}(t) = \sqrt{(k/a(t))^2 + m^2}   \label{2-7a}
\ee
This approximation is good in the limit that all time-derivatives of $a(t)$ smoothly approach $0$.
The two solutions in (\ref{2-7}) are linearly independent. Then the general solution of the second-order ordinary differential equation (\ref{2-5}) can be written in the adiabatic approximation as a linear combination
\bea
h_k(t ) &\sim & \alpha_k \, (\omega_{k}(t))^{-1/2} \exp ( - i \int^t \omega_{k}(t') \,dt') \cr
 & & \mbox{}+ \beta_k \, (\omega_{k}(t))^{-1/2} \exp ( + i \int^t \omega_{k}(t') \,dt'),
\label{2-8}
\eea
where $\alpha_k$ and $\beta_k$ are complex constants that must satisfy
\be
|\alpha_k|^2 - |\beta_k|^2 = 1 \label{2-9}
\ee
because of the conserved Wronskian or scalar product.
We will assume that $a(t)$ and all its time-derivatives are smooth and well-defined. 

In order to determine the expectation value of the number of  particles that are created as a result of a smooth expansion of the universe, we take $a(t)$ such that $\lim_{t \rightarrow -\infty} a(t) = a_1$, 
and $\lim_{t \rightarrow \infty} a(t) = a_2$, where $a(t)$ and all its time-derivatives exist and are well-defined and continuous, and $0 < a_1 < a_2$. 

We can take the solution of (\ref{2-5}) in the early-time Minkowski spacetime such that 
$|\alpha_k|^2 = 1$ and  $|\beta_k|^2 = 0$. Then the expression for the quantized field in (\ref{2-4a}) is the
standard expression for the quantized field in the early time Minkowski spacetime, for which $a(t)=a_1$.

In the late-time Minkowski spacetime, when one uses the late-time asymptotic form in (\ref{2-8}), and separates the positive frequency and negative frequency terms in the late-time expression for quantized field $\phi$ in (\ref{2-4a}), one finds that the field $\phi$ at late times can be rewritten in the form:

\be
\phi = \sum _{\vec k}\left\{ B_{\vec k} g_{\vec k}(\vec x, t)
+ B^\dagger{}_{\vec k} g^*{}_{\vec k}(\vec x, t)\right\}, \label{2-10a}
\ee
where
\be 
g_{\vec k} = (2V a_2{}^3)^{-1/2} e^{i\vec k\cdot\vec x}  \exp ( - i \int^t \omega_{k}(t') \,dt'). \label{2-10b}
\ee
Here $k^i = 2\pi n^i/L$ with $n^i$ an integer, and $k = \vert\vec k\vert$, and
\[ B_{\vec k} = \alpha_{k}\, A_{\vec k} + \beta_{k}\, A^{\dagger}{}_{-\vec k}{}.  \]
This transformation corresponds to the creation of particle-antiparticle pairs of total momentum $0$ from the vacuum, as required by conservation of momentum. This linear transformation is an example of a Bogoliubov transformation, as I explained in \cite{Parker1969} on page 1061, in connection with particle creation by expanding universes. Later, analogous Bogoliubov transformations involving sets of creation and annihilation operators were used by Fulling \cite{Fulling1973} to describe particles in the spacetime (Rindler spacetime) corresponding to a set of accelerating observers and by Hawking \cite{Hawking1974} to describe the distribution of particles created by a black hole.

In the expanding universe under consideration here, the expectation value of the number of particles created by the expanding universe from the early-time vacuum state $\vert 0 \rangle$, present at late times is
\be
\langle 0 \vert  B^{\dagger}{}_{\vec k} B_{\vec k} \vert 0 \rangle = \vert \beta_{k} \vert ^2
\label{2-10c}
\ee
(Recall that we are using the Heisenberg picture to describe the dynamics, so the state vector does not change with time.) Hence, if $\beta_{k}$ is non-zero, there is particle creation as a result of the expansion of the universe between the initial and final Minkowski spacetimes.  

Except in very special cases, quanta of the minimally-coupled scalar field are created by the expansion of the universe. From our earlier discussion of the quantized linearized-Einstein gravitational field in the Lifshitz gauge, the same should be true for the creation of gravitons by the expansion of the universe. For the isotropically expanding FLRW universes, the particles are created in pairs to conserve the 3-momentum. I also calculated the detailed probability distribution of the created particles as a function of momentum.  I used similar methods in \cite{Parker1975} to obtain the probability distribution of particles created by a Schwarzschild black hole.

I also considered the creation of particles if the initial state is not the Minkowski vacuum state, but instead has a non-zero density of minimally-coupled scalar particles present at early times. More generally, I considered a mixed state  described by a statistical density matrix operator $\rho_{stat}$ in the initial Minkowski spacetime. I showed that there would be stimulated creation of these bosonic particles by the expansion of the universe.
The details of the calculation are described in \cite{Parker1966}, and in \cite{Parker1969} on pp. 1063--1064.

Taking an isotropic distribution of particles initially, and an FLRW expansion,
the average number of particles present in mode $\vec{k}$ at late times in the final Minkowski spacetime is
\be
\langle N_{\vec k} \rangle_{late} = 
Tr [ \rho_{stat}  B^{\dagger}{}_{\vec k}  B_{\vec k} ]= 
   \langle N_{\vec k} \rangle_{early} + \vert \beta_{k} \vert^2 (1 + 2 \langle N_{\vec k} \rangle_{early})
 \label{2-11}
\ee
The stimulated creation of bosons is evident in the factor involving $\vert \beta_{k} \vert^2 (1 + 2 \langle N_{\vec k} \rangle_{early})$. The average number of these spin-0 bosons present at late times is increased by stimulated creation resulting from the presence of a non-zero average number of these particles at early times. I also showed in \cite{Parker1966}, and in \cite{Parker1971} that the opposite is true in the case of fermionic particles created by the expanding universe. 

But as we shall see in Section (\ref{exponential expansion}), when we discuss the rapid rate of increase of $\vert \beta_{k} \vert^2$ during the exponential stage of the expanding universe, the initial number of particles present in the coordinate volume $L^3$ can greatly influence the creation rate and the magnitude of the number of particles present at late times.

A further treatment of the stimulated creation of scalar gravitational perturbations during inflation is given by Ivan Agullo and me in \cite{Agullo-Parker2011-1,Agullo-Parker2011-2} where we discussed its possible effect on the CMB bispectrum. In \cite{Agullo-NavarroSalas-Parker2012} with J.~Navarro-Salas, we considered the CMB trispectrum.
}

\section{Exponentially Expanding FLRW Universe}
\label{exponential expansion}

{\textbf
Next we show how the above equations, (\ref{2-10c}) and (\ref{2-11}), for the expectation value of the number of minimally-coupled scalar particles created in each mode $\vec {k}$ determine the time-dependence of $\vert \beta_{k} \vert^2$ in an exponentially-expanding, spatially-flat, FLRW universe.

Let us start with a quantized minimally-coupled scalar field in an initial flat spacetime invariant under the 10-dimensional Poincar{\' e} group. Suppose that the quantized scalar field is in its vacuum state having the 10 Poincar{\' e} symmetries. Let us assume that our universe undergoes a spontaneous fluctuation (with $a(t)$, ${\dot a}(t)$, and ${\ddot a}(t)$ continuous) that eventually takes it to an exponentially-expanding, spatially-flat universe with $a(t) \propto \exp(H t)$, with $H$ constant. After a time, suppose the quantized scalar field in this exponentially expanding deSitter universe has settled into a state having the 10 deSitter symmetries. Because we are working in the Heisenberg picture, the state vector does not change with time, so it is the state $\vert 0 \rangle$ that is annihilated by the operator $A_{\vec k}$, the annihilation operator for the scalar quanta in the original early-time Minkowski spacetime. However, the state $\vert 0 \rangle$ is not annihilated by the operators $B_{\vec k}$ that annihilate particles in the late-time Minkowski spacetime. Recall that In the Heisenberg picture, the quantized field operator $\phi$ evolves with time, and the $B_{\vec k}$ in the late-time Minkowski spacetime are obtained from the field operator at late-times. The particles that are present in the {\em late-time} Minkowski spacetime are the {\em real} (as opposed to virtual) particles that were created during the exponential de-Sitter portion of the expansion and thus survived the gradual smooth slow-down of the expansion to approach the late-time Minkowski spacetime.  

We will show from this and the equation for particle creation that $\vert \beta_{\vec k} (t) \vert^2$ must be proportional to $\exp(H t)/ \exp(H t_1)$, where $t_1$ is approximately the time that deSitter symmetry was reached by the particles being created, and $t>t_1$. We have from (\ref{2-10c}), in the Heisenberg picture, for particle creation by the expanding universe from the Minkowski vacuum state in the initial Minkowski space that
\be
\langle 0 \vert  B^{\dagger}{}_{\vec k} B_{\vec k} \vert 0 \rangle = \vert \beta_{k} \vert ^2
\nonumber
\ee
Suppose that the expansion $a(t)$ approaches deSitter, with $a(t) \propto \exp(H t)$ smoothly, with at least $a(t)$, ${\dot a}(t)$, and ${\ddot a}(t)$ continuous. (This is so that the minimally-coupled scalar field equation is well-defined at all times.)  Here $k^i = 2\pi n^i/L$ with $n^i$ an integer, and $k = \vert\vec k\vert$, and
\[ B_{\vec k}(t_1) = \alpha_{k}(t_1)\, A_{\vec k} + \beta_{k}(t_1)\, A^{\dagger}{}_{-\vec k}{}.  \]
This transformation corresponds to the creation of pairs of total momentum $0$ from the vacuum, as required by conservation of momentum. We also have as a consequence of the field equation that, as in (\ref{2-9}),
\[  \vert \alpha_{k}(t_1) \vert^2 - \vert \beta_{k}(t_1) \vert^2 = 1, \]
which is the condition for the commutation relation of boson creation and annihilation operators to obey Bose-Einstein statistics, i.e., to satisfy the correct commutation relations.

After deSitter invariance has been reached, we expect that the number of particles present per unit {\em physical} volume ($L a(t)^3$) is constant for $t>t_1$. Hence, if $t > t_1$, it follows that
\[ \vert \beta_{k}(t) \vert^2 /a(t)^3  \]
is constant.
Therefore, during the period for which inflationary equilibrium has been reached, we must have
\[ \vert \beta_{k}(t_2) \vert^2 /a(t_2)^3  =  \vert \beta_{k}(t_1) \vert^2 /a(t_1)^3,  \]
which implies that 
\be
\vert \beta_{k}(t) \vert^2 \propto  (\exp(H t))^3, 
\label{2-12}
\ee
after inflationary equilibrium has been reached. 

This result in (\ref{2-12}) was confirmed numerically by Glenz and Parker \cite{Glenz-Parker2009} by means of an explicit numerical integration, in which we took an expansion scale factor $a(t)$ that started expanding from flat spacetime, eventually becoming an exponentially expanding inflationary universe that expanded for more than 50 e-foldings and then gradually approached a flat spacetime at late times.   We joined the three segments of the expansion such that $a(t)$, $\dot{a}(t)$, and  $\ddot{a}(t)$ were continuous at the two joining points. We used the known analytic solutions of the quantum field differential equations in each of the three segments of the expansion of the universe. As a check, we confirmed that (\ref{2-9}) was satisfied to at least 500 significant figures at all times throughout the expansion of the universe.

As an interesting aside, let me mention that the form of the parametrized scale factors $a(t)$ that we used in \cite{Glenz-Parker2009} to smoothly join the exponential inflationary expansion of the universe to Minkowski spaces at early and at late times are instances of the same four-parameter scale factor that I had used in a 1976 paper in Nature\cite{Parker1976} to demonstrate that an expanding universe can create free particles from vacuum, with the created particles having a thermal distribution in the absence of any thermalizing interactions.

The time-dependence of $\vert \beta_{k}(t) \vert^2$ in (\ref{2-12}) also follows in the case when an initial distribution of particles is present, as in (\ref{2-11}).
This is clear because the average number of created particles that is proportional to the average number, $\langle N_{\vec k} \rangle_{early}$, of particles already present at early times in (\ref{2-11}) is 
$(1+2\vert \beta_{k} \vert^2)\langle N_{\vec k} \rangle_{early} \approx
2\vert \beta_{k} \vert^2 \langle N_{\vec k} \rangle_{early}$,
so the argument given above leads in this case also to the same time-dependence, (\ref{2-12}), of 
$\vert \beta_{k} \vert^2$.  However, we see that the average number of particles created during the exponential expansion can be greatly increased by the stimulated creation of particles, if there were a significant average number $\langle N_{\vec k} \rangle_{early}$ of these bosonic particles already present in the physical expanding volume when the exponential stage of the expansion first began.
}

\end{document}